\documentclass[]{raa}            
\usepackage{graphicx,times}
\usepackage{natbib}

\providecommand{\bm}[1]{\mbox{\boldmath$#1$\unboldmath}}

\begin{document}

   \title{Kinematics of an untwisting solar jet in polar coronal hole observed by SDO/AIA}

 \volnopage{ {\bf 2011} Vol.\ {\bf } No. {\bf XX}, 000--000}
   \setcounter{page}{1}

   \author{H. Chen
      \inst{1,2}
 \and J. Zhang
      \inst{2}
   \and S. Ma
      \inst{1,3}   }
   \institute{College of Science, China University of
       Petroleum, Qingdao 266555, China; {\it hdchen@upc.edu.cn}\\
      \and
         Key Laboratory of Solar Activity, National Astronomical Observatories, Chinese Academy of Sciences,
      Beijing 100012, China;
                     \and
             Harvard-Smithsonian Center for Astrophysics, MA 02138, USA;
\vs \no
   {\small Received [year] [month] [day]; accepted [year] [month] [day] }
}

\abstract{Using the multi-wavelength data from the Atmospheric Imaging Assembly (AIA)
on board the Solar Dynamics Observatory (SDO) spacecraft,
we study a jet occurred in coronal hole near the northern pole of the Sun. The jet presented
distinct helical upward motion during ejection. By tracking six identified moving features (MFs)
in the jet, we found that the plasma moved at an approximately constant speed along the jet's
axis, meanwhile, they made a circular motion in the plane transverse to the axis. Inferred from
linear and trigonometric fittings to the axial and transverse heights of the six tracks,
the mean values of axial velocities, transverse velocities, angular speeds, rotation
periods, and rotation radiuses of the jet are 114 km s$^{-1}$, 136 km s$^{-1}$, 0.81\degr\ s$^{-1}$, 452 s,
and 9.8 $\times$ 10$^{3}$ km respectively.
As the MFs rose, the jet width at the corresponding height increased.
For the first time, we derived the height variation of the longitudinal magnetic field
strength in the jet from the assumption of magnetic flux conservation.
Our results indicate that, at the heights of 1 $\times$ 10$^{4}$ $\sim$ 7 $\times$ 10$^{4}$ km from jet base,
the flux density in the jet decreased from about 15 to 3 G as a function of B=0.5(R/R$_{\sun}$-1)$^{-0.84}$ (G).
A comparison was made with the other results in previous studies.
}

   \authorrunning{H. Chen, J. Zhang, \& S. Ma
   }            
   \titlerunning{Kinematics of an untwisting solar jet in polar coronal hole observed by SDO/AIA}  
   \maketitle

%
%
\section{Introduction}           
\label{sect:intro}
Solar jets are small-scale plasma ejections along straight or slightly curved coronal fields
(e.g., Shibata et al. 1994; Li et al. 1996; Chae et al. 1999).
They can be observed as emission in Ultraviolet (UV; e.g., Schmieder et al. 1988; Chen et al. 2008), Extreme-ultraviolet (EUV;
e.g., Schmahl 1981; Alexander \& Fletcher 1999; Kamio et al. 2007; Kim et al. 2007; Chifor et al. 2008a, 2008b; Kamio et al. 2009;
Yang et al. 2011a; Tian et al. 2011),
soft X-ray (SXR; e.g., Shibata et al. 1992; Zhang et al. 2000; Cirtain et al. 2007; Moore et al. 2011) and white light
(WL; e.g., Wang et al. 1998a; Liu et al. 2005b).
The detailed statistical properties of X-ray jets was studied by Shimojo et al. (1996) and more recently by Savcheva et al. (2007).
In morphology, surges are very similar to jets, but they appear as absorption features when observed on solar disk.
Sometimes, surges are observed to be associated with filament formation (e.g., Liu et al. 2005a), filament eruption
(e.g., Chen et al. 2009a; Guo et al. 2010; Moore et al. 2010; Hong et al. 2011) and even coronal mass ejections (CMEs,
e.g., Liu et al. 2005b; Jiang et al. 2008).
Generally speaking, surges and jets are the different appearances in different wavelengths of the same phenomenon.
In the following context, we use the term ``jets'' refers to both surges and jets.

Helical or twisted structures in jets have been reported by many authors (e.g., Dizer 1968; Shibata et al. 1992;
Canfield et al. 1996; Wilhelm et al. 2002; Jibben \& Canfield 2004; Jiang et al. 2007; Liu et al. 2009;
Shen et al. 2011; Liu et al. 2011; Curdt \& Tian 2011).
By using line of sight velocity field (e.g., Xu et al. 1984; Gu et al. 1996; Jibben \& Canfield 2004) and
stereoscopic (e.g., Patsourakos et al. 2008; Nistic\`{o} et al. 2009) observations,
some researchers confirmed that the rotation motions in some jets are real.
Xu et al. (1984) proposed a double-pole diffusion model to explain the rotating motion of a surge.
However, in consideration of the close relationship between jets and photospheric magnetic flux activities,
such as magnetic flux emergence, convergence, and cancelation etc (e.g., Roy 1973; Wang \& Shi 1993;
Shimojo et al. 1998; Zhang et al. 2000; Liu \& Kurokawa 2004; Chen et al. 2008),
more authors incline to think that the spinning of jets results from the relaxation of magnetic twist, which occurs
when twisted photospheric magnetic loop reconnects with ambient open fields (e.g., Shibata \& Uchida 1986; Shibata et al. 1994;
Canfield et al. 1996; Patsourakos et al. 2008; Nistic\`{o} et al. 2009; Kamio et al. 2010; He et al. 2010).
Recently, three-dimensional simulations by Pariat et al. (2009) show that
high-level magnetic stress due to twisting motion can lead to an explosive release of energy via
reconnection, which will produce massive, high-speed jets driven by nonlinear Alfv\'{e}n wave.
And if the stress is constantly applied at the photospheric boundary, this mechanism would generate
recurrent untwisting quasi-homologous jets (e.g., Pariat et al. 2010; Asai et al. 2001; Chen et al. 2008; Yang et al. 2011b).
More recently, the simulations by D\'{i}az et al. (2011) indicate that the speed of the flow along the field lines of twisted magnetic
flux tubes may be super-Alfv\'{e}nic and the twisted tube is subject to the kink instability, which could explain the
behaviour of super-Alfv\'{e}nic jets and the disruption of some observed jets.

As mentioned above, so far, the main observational methods to investigate the spining of jets are focused on
the analysis of line of sight velocity field and stereoscopic observations, or taking advantage of the
technique of time-distance analysis (e.g., Liu et al. 2009).
Lower temporal and spatial resolutions of these observations or the limitation of the technique used in these studies
can not make the exact kinematics of jet clear.
For example, the tracks or stripes in the time-distance slit images can not stand for the real motion of jet plasma along the
slit direction due to the perpendicular velocity.
The Atmospheric Imaging Assembly (AIA; Lemen et al. 2011) on the Solar Dynamics Observatory (SDO; Schwer et al. 2002)
images the solar atmosphere in 10 wavelengths with 12 s high temporal resolution.
The instrument observes solar plasma from photosphere to low corona with a full-disk field of view and the pixel size is about 0.6\arcsec.
High-resolution AIA intensity images can reveal the fine structures in jets, which provides us an opportunity to
track the motions of some moving features (MFs) in jets.
Using this new method, we study the detailed kinematics of one AIA 304 \AA\ jet, which has been investigated by
Shen et al. (2011) mainly using the technique of time-distance slit images.
One aim of this paper is to compare the results from the two different methods.

In addition, the measurement of coronal magnetic field strength is a long-standing unresolved problem in
solar physics (e.g., West et al. 2011).
Due to thermal broadening and polarization effect, usual methods for measuring coronal flux density,
such as Zeeman splitting of spectral lines and Hanle effect, become complicated.
Focusing on the stronger active region fields, Lin et al. (2004) measured the magnetic flux density
100$\arcsec$ ($\sim$7 $\times$ 10$^{4}$ km ) above an active region to be 4 G, which is smaller
than the results (10$\sim$33 G) presented by an earlier work of Lin et al. (2000).
Some indirect methods, such as photospheric extrapolation techniques (e.g., Wang \& Sheeley 1992;
Liu \& Lin 2008), radio techniques (e.g., Ramesh et al. 2010), and coronal seismology
(e.g., Uchida 1970; Roberts et al. 1984; Chen et al. 2011; West et al. 2011; Gopalswamy \& Yashiro 2011),
are applied to estimate the coronal magnetic field.
In this study, in combination with the observations of the studied jet, we try to provide a new technique
to estimate the magnetic field of the higher part of the jet.
This would give valuable insight into the coronal magnetic field structure.

In the next section, we briefly describe the observations and data used in our study. This is followed by
a detailed study of the kinematics of the jet and an estimation of the longitudinal magnetic flux density in the jet.
Finally, we give the summaries and discussions.

\begin{figure}[!ht]
   \centering
   \includegraphics[width=12.0cm, angle=0]{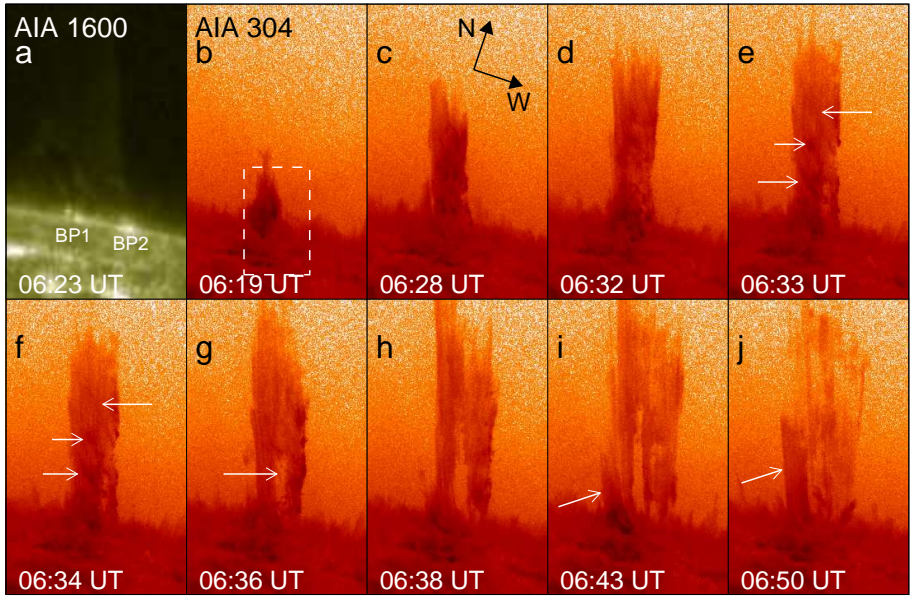}
   \begin{minipage}[]{120mm}
   \caption{ Panel a: AIA 1600 \AA\ image displaying the brightening patches (BPs) at the base of jet.
   Panels b-j: negative AIA 304 \AA~images showing the detailed evolution of the jet.
   The images have been clockwisely rotated by 18$\degr$ from
   the northern pole of the Sun, which is the same for all AIA images in Fig. 2 and 3.
   The field of view (FOV) of 304 \AA\ images is 132\arcsec $\times$ 216\arcsec.
    The dashed box in panel b indicates the FOV of panel a, which is 48\arcsec $\times$ 79\arcsec.}
   \end{minipage}
   \label{Fig1}
   \end{figure}

\section{Data and Observations}
\label{sect:Obs}
On 2010 August 21, a jet occurred at the northeast limb (E0N81) of the Sun (Shen et al. 2011).
The observation from Extreme UltraViolet Imager (EUVI; Wuelser et al. 2004) of Sun Earth Connection Coronal
and Heliospheric Investigation (SECCHI; Howard et al. 2008) on board the spacecraft B of Solar Terrestrial
Relations Observatory (STEREO; Kaiser et al. 2008)
indicates that the jet is rooted in coronal hole (e.g., Zhang et al. 2007) near the northern pole of the Sun.
The detailed evolution of the jet was observed by the AIA on SDO, which provides multiple simultaneous high-resolution full-disk images
up to 0.5R$_{\sun}$  above the solar limb with 1.2 arcsec spatial resolution and 12 s cadence.
All the ten bandpasses have been employed in the observations of this jet activity.
In this paper, we mainly used the channels centered at 304 \AA, 1600 \AA, 171 \AA,
193 \AA, and 211 \AA\ (Level 1.5 images) with the temperature responses peak at 0.05 MK, 0.1 MK, 0.6 MK, 1.5 MK (also 20 MK),
and 2.0 MK, respectively (Lemen et al. 2011).
We did not perform any de-rotation since the rotation effect will not influence our results significantly.

\section{Results}
\label{sect: data}

\subsection{General Evolution of the Jet}

Figure~1 shows the morphology and general evolution of the jet at AIA 304\AA\ (reversed color table).
Since the projected direction of the jet's axis is about 18\degr anti-clockwise from
the northern pole of the Sun, all the AIA images in this paper have been rotated the same angle clockwisely
for better showing.
According to the AIA observations, the jet took place at about 06:07 UT, when a brightening patch BP1 (see panel a of
Fig. 1) firstly began to appear at one (eastern) side of the root and gradually evolved into a apparent inverted ``Y'' structure
in 171 \AA\ images.
Since then, dense plasma began to flow out from BP1 and expanded westwardly.
From AIA 1600 \AA\ images, we can see that another brightening patch BP2 (in panel a of Fig. 1)
appeared at the opposite (western) side of the base region at about 06:18 UT and peaked at 06:23 UT.
In combination with 304 \AA\ observations, it seems that the main mass of jet was ejected from above
BP2 rather than BP1 after BP2 appeared. We consider that this location change of jet footpoint
has a close association with the magnetic reconnection occurred at the jet base.

   \begin{figure}[!ht]
   \centering
    \includegraphics[width=12.0cm, angle=0]{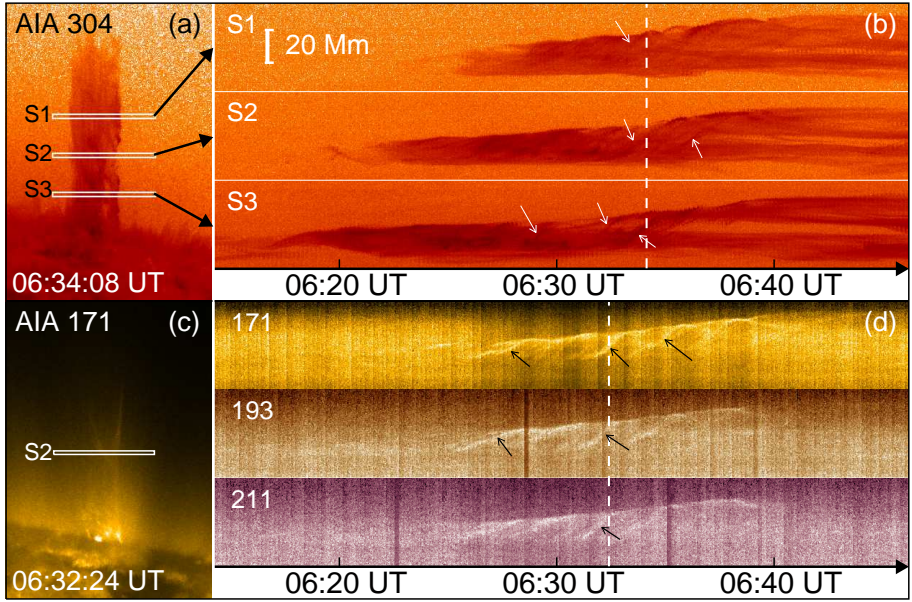}
   \begin{minipage}[]{120mm}
   \caption{Panels a and c are AIA 304 \AA\ and 171 \AA\ intensity images, respectively. They have a same FOV of
   150\arcsec $\times$ 216\arcsec.
   The white narrow boxes indicate the positions of the slits S1-S3, which have a FOV of 74\arcsec $\times$ 4\arcsec.
   Panel b: slit images from AIA 304 \AA\ channel along S1-S3, respectively.
   Panel d: slit images along S2 from AIA 171 \AA, 193 \AA, and 211 \AA\ channels, respectively.
   The two white dashed lines indicate the time 06:34:08 and 06:32:24 UT when the AIA 304 \AA\
   and 171 \AA\ intensity images (in panels a and c, respectively) were recorded. The arrows in panels b and d point to some
   stripes, which indicate the transverse motions of the plasma across the jet.
   }
   \end{minipage}
   \label{Fig2}
   \end{figure}

As the plasma was ejected outwards, the jet also spun clockwise as viewed from its footpoints.
Because of the movements along both axial and transverse (rotation) direction, the jet appeared as upward helical structures.
Some fine twisted threads with a mean width of a few arcsecs can be identified clearly, which are indicated by the white
arrows in panels e and f of Fig. 1. According to the charity definition of jet in Jibben and Canfield (2004),
the jet we studied here is a right-hand jet. As time went on, these threads
gradually unwound, and one big bifurcation (indicated by the arrow in panel g of Fig. 1)
appeared at about 06:35 UT from the bottom and spread upward along the body of the jet.
At about 06:45 UT, after reaching a maximum height of 17.9 $\times$ 10$^4$ km,
the material began to fall back along almost the axial direction without any transverse motion.

Interestingly, we note that another jet (indicated by the arrows in panels i and j of
Fig. 1) occurred at about 06:42 UT, when the first has not completely disappeared yet.
Its feet are very close to the feet of the first jet, which means they likely originated
from the same source region. However, the ejection directions of the two jets
are not very consistent with each other. Similar phenomenon has also been reported
by Chen et al. (2008).
In recent simulation of Pariat et al. (2010), their results show that the drifting directions are different for recurrent jets,
even if the underlying magnetic system and the driving motion remain constant.

\begin{figure}[!ht]
   \centering
   \includegraphics[width=10.0cm, angle=0]{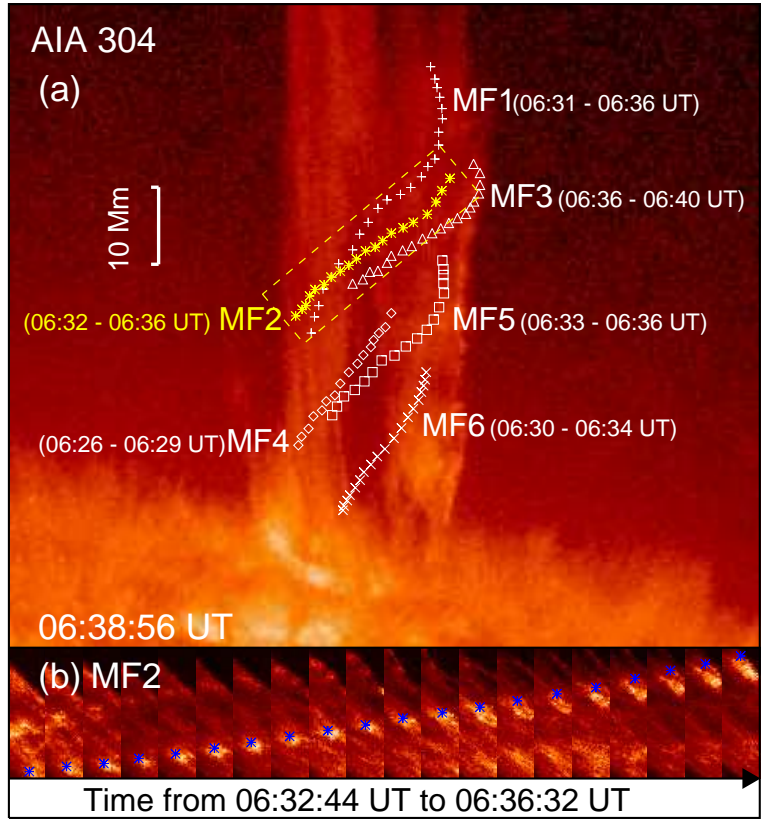}
   \begin{minipage}[]{100mm}
   \caption{Panel a: one AIA 304 \AA\ image overlaid with the tracks of MF1-MF6 (plus, asterisk, triangle, diamond,
   square, and $\times$, respectively).
  The time in the parentheses are the start and end tracking time of the corresponding MFs.
  The FOV of panel a is 168\arcsec $\times$ 144\arcsec.
   The yellow dashed box indicates the FOV of the slit image in panel b, which is about 14\arcsec $\times$ 52\arcsec.
   Panel b: slit images showing MF2. The blue asterisks indicate the positions of MF2 at different time.
    } \end{minipage}
   \label{Fig3}
   \end{figure}

\subsection{Helical Upward Motion}
A remarkable character of this jet is its distinct transverse rotating motion.
In Figure~2, we show this transverse motion in detail.
Two AIA intensity images in 304 \AA\ and 171 \AA\ wavelengths are given in panels a and c of Fig. 2, respectively.
The three white narrow boxes (74\arcsec $\times$ 4\arcsec) in the 304 \AA\ images mark three slits S1-S3 from top to bottom,
which are perpendicular to the jet axis. The heights of S1-S3 from the jet base are about 2.36 $\times$ 10$^4$,
4.43 $\times$ 10$^4$, and 6.50 $\times$ 10$^4$ km, respectively.
In panel b of Fig. 2, we display the time-distance diagrams at 304 \AA\ along slits S1, S2 and S3
from top to bottom, respectively.
As shown in these time-distance diagrams, it can be seen that there are many stripe structures,
which indicate the transverse motion of the plasma in the jet.
In total, fifteen stripes can be clearly identified in panel b, among which several typical ones are indicated by
the white arrows.

We performed linear fittings to all the fifteen time-distance tracks, and found that the transverse velocities
of these features range from 70 to 200 km s$^{-1}$ with a mean value of 134 km s$^{-1}$.
Using the same method, a more detailed investigation on the transverse motion of this jet has been done by Shen et al. (2011).
According to their results, the total average transverse velocity of the jet is 123 km s$^{-1}$.
In this paper, using the observations from other bandpasses, we further extended this study.
In panel d of Fig. 2, the time-distance diagrams from 171 \AA, 193 \AA, and 211 \AA\ intensity
images are plotted from top to bottom, respectively.
Because the transverse rotation is not very clear in 171 \AA, 193 \AA, and 211 \AA\
observations along S1 and S3, the time-distance diagrams at slit S2 are shown alone.
Similarly, from the slit images in panel d, the transverse helical motion features can be identified clearly,
some of which are marked by the black arrows. Furthermore, in morphology, these transverse motion features observed
in the three EUV wavelengths are very similar.
According to the linear fitting results, the mean transverse velocity of these features is around 140 km s$^{-1}$,
which is similar to the result from 304 \AA\ slit images.

\begin{figure}[!ht]
   \centering
   \includegraphics[width=10.0cm, angle=0]{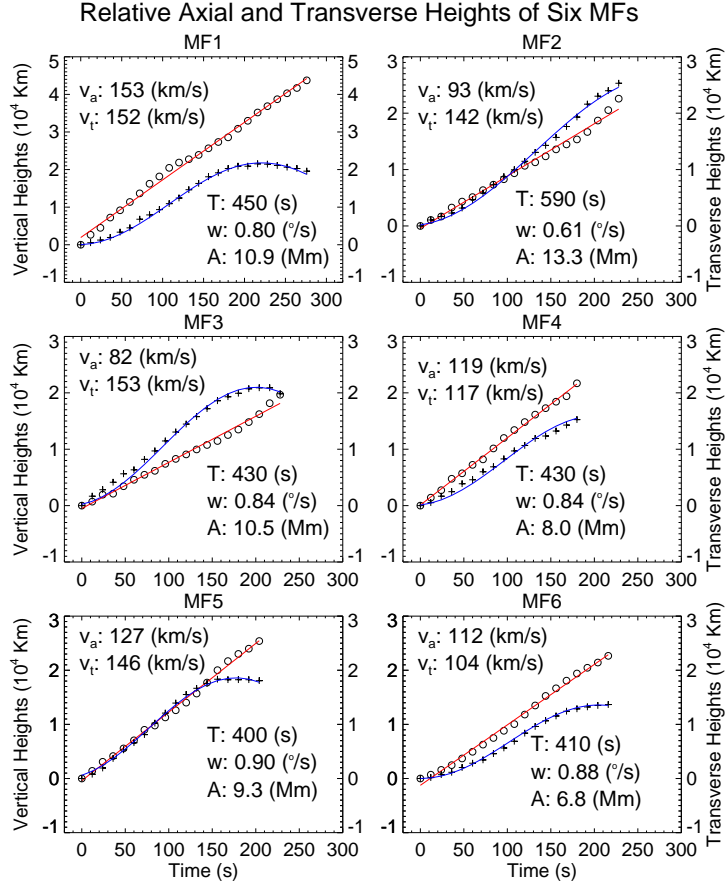}
   \begin{minipage}[]{100mm}
   \caption{Time variations of the relative axial (circle) and transverse (plus) heights of MF1-MF6.
   The red and blue solid lines are the results of linear and trigonometric fittings to the axial
   and transverse heights, respectively. Here, v$_{a}$, v$_{t}$, T, $\omega$, and A represent the axial velocity,
   transverse velocity, rotation period, angular speed and rotation radius of the MFs, respectively.
} \end{minipage}
   \label{Fig4}
   \end{figure}

   \begin{figure}[!ht]
   \centering
    \includegraphics[width=12.0cm, angle=0]{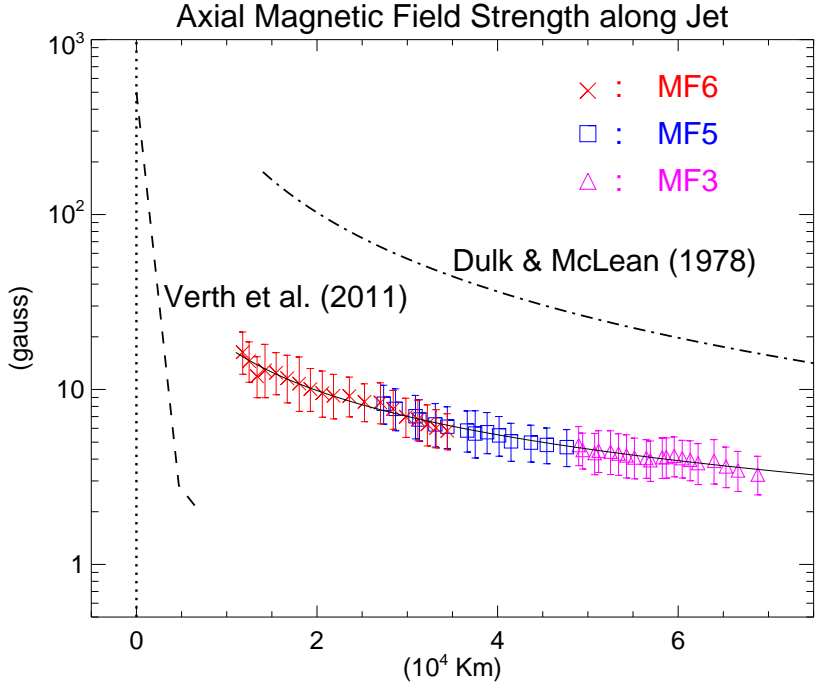}
   \begin{minipage}[]{100mm}
   \caption{Height variation of axial magnetic field strength in the jet. The $\times$ (red), square (blue),
    and triangle (purple) symbols correspond to the data derived from the tracks of MF6, MF5, and MF3, respectively.
    The dashed and dash-dotted lines are the results from Verth et al. (2011) and Dulk \& McLean (1978), respectively.
    The solid line is the fit to our observational data. The dotted line indicates the position of solar surface.
    } \end{minipage}
   \label{Fig2}
   \end{figure}

To reveal the kinematics of the jet more clearly, we tracked six moving features (MFs) that
can be clearly identified in the jet. Considering the freezing-in effect of plasma-magnetic
field coupling, we assume that each MF moved along the same magnetic field line during ejection.
In panel a of Figure 3, one AIA 304 \AA\ image at 06:38:56 UT
is overlaid with the tracks of MF1-MF6 (marked with plus, asterisk, triangle, diamond, square, and $\times$  symbols, respectively).
The start and ending tracking time of each MF are shown in the corresponding parentheses.
It can be seen from Fig.3 that most of MFs' tracks appear like helical lines, which indicates the MFs
made helical upward motions in the jet.As an example, we show the evolution of MF2 in panel b of Fig.3
The blue asterisks in panel b indicate the positions of MF2 at different time.

In Figure 4, we plot the time profiles of axial (circle) and transverse (plus) heights of the six MFs.
Note that the time and heights in each panel are the values relative to the initial time and heights of respective MF.
According to the different distributions of the axial and transverse heights, we can see that all the MFs seemed to move
at a approximately constant speed along the jet's axis and make a circular motion across the jet at the meantime.
We performed linear (red solid line) and trigonometric (blue solid line) fittings to the
axial and transverse heights of each MF, respectively. It can be found that the observational data
are fitted very well, which provide further evidence of the helical motion of the jet.
The axial velocity (v$_{a}$), transverse velocity (v$_{t}$), angular speed ($\omega$), rotation period (T),
and rotation radius (A) of each MF are derived from the linear and trigonometric fitting results
and shown in the corresponding panel of Fig. 4.
The mean values of v$_{a}$, v$_{t}$, $\omega$, T and A are 114 km s$^{-1}$, 136 km s$^{-1}$, 0.81\degr s$^{-1}$
(or 14.1 $\times$ 10$^{-3}$ rad s$^{-1}$), 452 s, and 9.8 $\times$ 10$^{3}$ km , respectively.
In comparison with the results from Shen et al. (2011), we found most of the results are similar except
the mean axial velocities, which are 114 and 171 km s$^{-1}$ in our and their studies, respectively.
This difference maybe results from the limitation of our MFs sample number.

In additon, from the fitting results in our study, there seems to be no obvious correlation between v$_{a}$ and v$_{t}$.
For example, the transverse velocities of MF1, MF2 and MF3 are about 150 km s$^{-1}$; however, the
axial velocity (153 km s$^{-1}$) of MF1 is much bigger than those (93 and 82 km s$^{-1}$) of MF2 and MF3.
Of course, for more accurate statistical relationship between the axial and transverse velocity of jet plasma,
further study based on more samples is needed.
Using these results, we roughly estimate the twist spreading into the outer
corona during ejection, which may be restored in the photospheric flux rope before.
The AIA 304 \AA\ movie shows that the total spinning period of the jet is approximately from 06:16 UT
to 06:42 UT ($\sim$26 minutes). Assuming that the jet made an uniform circular motion, the
total twist restored can be yielded by dividing the total spinning time by the mean rotation period (452 s),
which is about 3.6 turns. In contrast to the result presented by Shen et al. (2011), ours is bigger..

\subsection{Axial Magnetic Field Strength in the Jet}
As mentioned at the beginning, in this study, we try to provide a new method to estimate
the longitudinal magnetic field in the jet.
Our basic idea is that assuming the jet plasma flows in the same flux tube during ejection, the magnetic flux
across the transverse section of the jet would keep constant, i.e.
\begin{equation}
B_{o}S_{o}=BS
\end{equation}
So, if we can determine the photospheric magnetic field strength (B$_{o}$), transverse area (S$_{o}$) of
the flux tube at the jet base and the transverse area (S) of the flux tube at a certain height, then
the axial magnetic field strength (B) at the corresponding height can be derived from equation (1).

We describe the determinations of B$_{o}$, S$_{o}$, and S as below.
First, we approximately think the flux tube, i.e. the channel along which the jet material flows, as a
axisymmetric cylinder with a increasing radius (r). So, the transverse area S$_{o}$ and S can be represented by
$\pi$$r_{o}^{2}$ and $\pi$$r^{2}$, respectively.
Due to dispersion, it is difficult to measure the jet's radius (r$_{o}$) at the base directly.
Thus, r$_{o}$ is estimated by the size of the brightening patch BP2 appeared at the jet base in AIA 1600 \AA\
image, which implies r$_{o}$ is about 2.6 $\times$ 10$^{3}$ km.
By tracking the axial heights of the MFs at different time, we measured the width of jet at the corresponding
heights, which is twice the size of r.
As for B$_{o}$, since the jet occurred at solar rim and there is no available photospheric magnetic field data,
here, we take the mean photospheric flux density value ($\sim$500 G) from Chen et al. (2008) as B$_{o}$.
Considering the similar spatial and temporal scales of jets in this study and Chen et al. (2008), the used value (500 G)
of B$_{o}$ should be reasonable.

Using the corresponding data from the tracks of MFs, the axial magnetic field strength (B) along the
jet were derived from equation (1) and its variation with height (from the base of jet) is shown
in Figure~5. Note that we only used the data from three tracks of MF3 (purple triangle), MF5 (blue square),
and MF6 (red $\times$). This is because not only the axial heights of the three MFs but also their
evolution time have better succession than the other's (see Fig. 3).
The errors of B mainly result from the uncertainty in measurements of r$_{o}$ and r and increase
as the heights decline. From Fig. 5, it can be seen that B decreases with increasing height.
Especially, it falls quickly at the lower height. According to our results, B decreases half
(from about 15$\pm$4 to 7$\pm$2 G) from the height of 1.1 $\times$ 10$^{4}$ km to 2.8 $\times$ 10$^{4}$ km, with
a mean drop rate of 4 G per 10$^{4}$ km.
As the height keeps increasing, B gradually declines to about 3$\pm$1 G at the height of 7 $\times$ 10$^{4}$ km.
From the heights of 2.8 $\times$ 10$^{4}$ to 7 $\times$ 10$^{4}$ km, the mean decline rate of B is about 1 G per 10$^{4}$ km,
which is only one fourth of that below 2.8 $\times$ 10$^{4}$ km.

Taking advantage of the same relationship of the magnetic flux density with the width of flux tube, i.e. B $\sim$ 1/r$^2$,
Verth et al. (2011) studied the magnetic field strength along a solar spicule.
Their results (dashed line) are shown in Fig. 5. Obviously, the flux density derived in
Verth et al. (2011) drops more quickly at the typical heights (from photosphere to 7 $\times$ 10$^{3}$ km) of a spicule.
As a comparison, we also show the results (dash-dotted line in Fig. 5) from the empirical active region
magnetic field model (Dulk \& McLean 1978), which is given by B=0.5(R/R$_{\sun}$-1)$^{-1.5}$ G.
On the whole, the field strength values from the model are about six times of our results. In consideration of the
different magnetic field structures between above active region and in the coronal hole and the certain
errors of measurements, we think our results are reasonable.
By revising slightly to the empirical formula presented by Dulk \& McLean (1978), we found
a new formula
\begin{equation}
B=0.5(R/R_{\sun}-1)^{-0.84}~~G
\end{equation}
fits our observations well, wherein, R is the distance from the solar center.
In Fig. 5, the fitting results are indicated by the black solid line across the colored symbols.

\section{SUMMARY}
In this paper, we present a detailed study of a jet which showed a distinct transverse rotating motion during its ejection.
The observational results appear to be consistent with an untwisting model of magnetic reconnection (e.g., Shibata \& Uchida 1986;
Pariat et al. 2009; 2010).
By tracking six identified features moving helically in the jet, we found that the jet plasma moved at an approximately
constant velocity along the axial direction and made a circular motion in the plane perpendicular to the jet axis.
we derived the axial velocity, transverse velocity, angular speed, rotation period and rotation radius for each MF.
Their mean values are 114 km s$^{-1}$, 136 km s$^{-1}$, 0.81\degr s$^{-1}$, 452 s, and 9.8 $\times$ 10$^{3}$ km ,
respectively. By comparison with the other study using different method (Shen et al. 2011),
we found most of the results are similar. For more accurate kinematics of jet plasma,
a more extensive statistical investigation work is expected in the future.

On assumption of the magnetic flux conservation in the same flux tube, we made an estimation
of the field strength of the jet occurred in the polar coronal hole.
Our results show that the longitudinal flux density of the jet at the heights
of 1 $\times$ 10$^{4}$ $\sim$ 7 $\times$ 10$^{4}$ km from solar surface,
decreased from about 15 to 3 G. Comparing with the result from Dulk \& McLean (1978), a new formula
of B=0.5(R/R$_{\sun}$-1)$^{-0.84}$ (G) fits our estimated data well.
It should be noted that the B$_{o}$ used in our study is just an estimated value, which maybe
leads to a major error of the absolute value of B. However, it would not significantly affect the height variation of B.
On the other hand, since almost all of the direct (Lin et al. 2000; Lin et al. 2004)
or indirect (Cho et al. 2007; Ramesh et al. 2010; West et al. 2011) measurements
of coronal fields strength in previous studies are mainly focused on the stronger fields above or at least emanate from
major active regions,
our results could offer helpful information about the magnetic field structures above mini active region newly-emerged
in coronal hole.

The formation of the moving features (MFs) in the jet is also an interesting question.
Similar features in jets can be seen in some other observations (e.g., Jiang et al. 2007;
Liu et al. 2009; Liu et al. 2011). Although we know that both local density enhancement and
temperature enhancement might be responsible for the existence of the MFs.
However, why and how the local density or temperature enhancement takes place are still
obscure at present time. We suggest that three possible mechanisms might contribute to the formation of MFs.
First, the formation of the MFs might be associated with the successive occurrence of magnetic reconnection at the jet base.
Second, the intrinsic (sausage or kink) instability in the mass flow as described in Chen et al. (2009b) and D\'{i}az et al.(2011)
might be another possible candidate.
In morphology, the MFs are similar with the plasma blobs observed in coronal streamers (e.g., Sheeley et al. 1997;
Wang et al. 1998b; Wang et al. 2000; Song et al. 2009).
The simulations by Chen et al.(2009b) reveal that the sausage-kink instability of coronal streamers
may lead to the formation of the plasma blobs. At this point, we think that the production mechanism of MFs
in jets may be similar to that of the plasma blobs. In addition, the simulations of D\'{i}az et al.(2011) support
that the kink instability in the mass flow can result in the disruption observed in solar jets.
At last, the intrinsic unevenness of plasma density in the photospheric twisted flux tube might
also be a possible formation mechanism of the MFs in solar jets.



\normalem
\begin{acknowledgements}
The authors sincerely thank the anonymous referee for very helpful and constructive
comments that improved this paper.
We are grateful to all the members of the Solar Magnetism and Activity group of
National Astronomical Observatory of CAS for invaluable help.
We acknowledge the AIA team for the easy access to calibrated data. The AIA
data are courtesy of SDO (NASA) and the AIA consortium.
This work was supported by the National Natural Science Foundation of China
(11103090, G11025315, 40890161, 10921303, 40825014, and 40890162), the CAS project KJCX2-YW-T04, the
National Basic Research Program of China under grant G2011CB811403, and
Shandong Province Natural Science Foundation (ZR2011AQ009).
\end{acknowledgements}



\label{lastpage}

\end{document}